\documentclass[%
 reprint,
 superscriptaddress,
 amsmath,amssymb,
 aps,
]{revtex4-1}

\usepackage[colorlinks, citecolor=red]{hyperref}
\usepackage{graphicx}
\usepackage{dcolumn}
\usepackage{bm}

\usepackage{multirow}
\usepackage{listings}
\usepackage{color}
\usepackage{subfigure}
\usepackage{amsmath}
\usepackage{bbm}
\usepackage{makecell}
\usepackage{ntheorem}
\usepackage[braket, qm]{qcircuit}
\usepackage{verbatim}
\usepackage{float}

\definecolor{dkgreen}{rgb}{0,0.6,0}
\definecolor{gray}{rgb}{0.5,0.5,0.5}
\definecolor{mauve}{rgb}{0.58,0,0.82}

\usepackage{algpseudocode}
\usepackage{algorithm}

\algnewcommand\algorithmicforeach{\textbf{for each}}
\algdef{S}[FOR]{ForEach}[1]{\algorithmicforeach\ #1\ \algorithmicdo}

\begin{document}

\title{Origin Pilot: a Quantum Operating System for Effecient Usage of Quantum Resources}
\author{Weicheng Kong}
\affiliation{Origin Quantum Computing Company Limited, Hefei 230026, China}
\affiliation{CAS Key Laboratory of Quantum Information (University of Science and Technology of China), Hefei 230026, China}
\author{Junchao Wang}
 \affiliation{CAS Key Laboratory of Quantum Information (University of Science and Technology of China), Hefei 230026, China}
 \affiliation{State Key Laboratory of Mathematical Engineering and Advanced Computing, Zhengzhou 450002,China}
\author{Yongjian Han}
 \affiliation{CAS Key Laboratory of Quantum Information (University of Science and Technology of China), Hefei 230026, China}
\author{Yuchun Wu}
 \affiliation{CAS Key Laboratory of Quantum Information (University of Science and Technology of China), Hefei 230026, China}
\author{Yu Zhang}
 \affiliation{School of Computer Science and Technology, University of Science and Technology of China, Hefei 230027, China}
\author{Menghan Dou}
 \affiliation{Origin Quantum Computing Company Limited, Hefei 230026, China}
\author{Yuan Fang}
 \affiliation{Origin Quantum Computing Company Limited, Hefei 230026, China}
\author{Guoping Guo}
 \affiliation{Origin Quantum Computing Company Limited, Hefei 230026, China}
 \affiliation{CAS Key Laboratory of Quantum Information (University of Science and Technology of China), Hefei 230026, China}

\begin{abstract}

The operating system is designed to manage the hardware and software resources of a computer. With the development of quantum computing, the management of quantum resources and cooperation between quantum systems and other computing resources (e.g. CPU, GPU and FPGA etc.) become the key challenge for the application of quantum computing to solve real world problems. In this paper we propose a quantum operating system, Origin Pilot. Origin Pilot includes the module of quantum task scheduling, quantum resource management, quantum program compilation and qubits' automatic calibration. With these modules, Origin Pilot can manage the quantum computing resources and solve the multi-quantum processor scheduling problem. It can also allow the parallel execution of multiple quantum programs and calibrate the quantum resource effectively. Thus, the performance of resources is guaranteed and the resource utilization is improved. By comparing the results with and without Origin Pilot, we evaluate the impact on a quantum circuit's fidelity of qubits mapping algorithm. We also evaluate the effectiveness of automatic calibration and parallel execution of multi-quantum processors. Finally, Origin Pilot can be easily customized for hybrid computing resources.    

\textbf{Keywords:} Origin Pilot; Quantum Operating System; Quantum Computing; Quantum Processor

\end{abstract}


\maketitle


\section{Introduction}

The quantum computing is a new computing paradigm based on the combination of quantum mechanics and computer science\cite{2002Quantum}. It provides enormous parallel computing power and storage which exceeds all classic computing technologies taking advantage of the quantum superposition and entanglement\cite{1996A}\cite{2009Quantum}. The qubits serves as the basic unit in a quantum computer. Compared with classic computing, the quantum computing can achieve exponential speedup in decryption\cite{1994Algorithms}, quantum chemistry\cite{2013A}, finance\cite{2018Identifying} and machine learning\cite{0Quantum}\cite{Petruccione2015An} etc. 

There are different physical approaches for implementing a quantum computer, such as the semiconductor spin\cite{Li2003An}, superconducting\cite{2012On}, trapped-ion\cite{H2008Quantum}\cite{2002Architecture} and optical systems\cite{2009Optical}\cite{2007Optical} etc. The quantum computer can be applied in its early stages with the improvement of material and manufacturing, optimization of environmental noise, control and electronic system development, the advancement in control architecture and basic quantum software\cite{2021TEQUILA}. Researchers also demonstrate the quantum advantage in recent two years. In 2019, Google proved that the Sycamore quantum processor with 53 qubits can exceed the most powerful super computers in random circuit sampling problem\cite{Arute_2019}. IBM published the quantum computing service in cloud based on the engineering advancement. Apart from IBM, there are other quantum computing service providers with real quantum computing backend, such as D-Wave, Google, Rigetti, Quantum Inspire and Origin Quantum etc.

However, with the increasing demand on quantum computing, how to effectively manage quantum computing infrastructures\cite{2010Computational} and use quantum computing resources\cite{2019Quantum} becomes one of the key problem. In 2015, Henry et al. introduces three quantum computing hardware architectures including quantum FPGA, quantum x86 system and quantum distributed computing system\cite{10.1145/3102980.3102993}. In 2020, Reid et al. propose a multi-programming approach that can execute multiple circuits in parallel by analyzing the dependency between circuits\cite{2020A}. There are two commercial companies publishing their quantum operating systems. Deltaflow.OS allows the same quantum circuit executed on different types of quantum computing hardware, which can allow quantum application developers focus more on the software and application itself. Parity OS can optimize a quantum circuit with the assistance of a quantum compiler which can further be compatible with a specific quantum processor.

Existing works try to optimize the performance of a quantum computer from different perspectives. Based on existing works, we find there are two problems which need to be solved:

1) Multiple quantum processors' scheduling. Current quantum cloud system only allows users to use only a single quantum processor at a time. Once the quantum processor is assigned to a user, it is fully occupied by the user. Other users cannot access the resources until the assigned user release the quantum processor. When the quantum computing service allows users to choose the quantum processors, the queuing time will far exceeding the execution time especially for the quantum processor with better performance. Although other quantum processors can meet the computing's requirement, the quantum task still needs to wait. Such situation can lead to the resource under utilization. The main reason is that there is no automatic resource allocation and scheduling for multiple quantum tasks. Thus, in this paper we design an algorithm to allocate quantum resources based on the requirements of the submitted quantum tasks. 

2) automatically optimize the quality of a qubit. The qubit can be easily disturbed by environment, which makes the qubits' performance fluctuate. The quantum gates' fidelity will decrease if the qubits' are not properly calibrated. An existing solution is to calibrate all the qubits when the performance deteriorates. However, the existing calibration approach has the following drawbacks:  a quantum circuit's fidelity cannot be guaranteed before calibration; the quantum processors cannot work during the long period of calibration. The Optimus is an automatic 
calibration system developed by Google. The Optimus can traverse all the qubits' state and deal with the ``bad" qubits in real time. However, they don't explicitly determine the state of qubits which are not calibrated. Moreover, they don't consider the situation that the calibration is conducted while there are other general quantum tasks.

To compensate the above problems, we propose Origin Pilot, which is a quantum operating system that can effectively use quantum resources. We implement four services to tackle the above problems, including quantum task scheduling, quantum resource automatic calibration, quantum circuit compiling and quantum resource management. 

The main contribution of Origin Pilot includes:

1) Origin Pilot can calibrate a single or multiple qubits online without interrupting other quantum circuits;

2) Origin Pilot can allocate quantum resources for quantum circuits according to the state of the quantum resources and the requirements of quantum circuits;

3) Origin Pilot can reduce the decoherent noise with dynamic decoupling in qubits. Thus, Origin Pilot allows multiple circuits executing on the same quantum processor. The average completion time for quantum circuits can be greatly reduced.

The rest of this paper is organized as follows: In Section 2 we introduce the basic concepts of quantum computing and quantum operating system; In Section 3 we propose the overall architecture and workflow of Origin Pilot; In Section 4 we describe the solution to the multi-quantum processor load balancing, multi-quantum program parallel computing and automatic calibration of qubits; In Section 5 we analyze the experimental results of Origin Pilot; In Section 6 we conclude this paper and propose the future works.

\section{Preliminary Knowledge}

\subsection{Quantum Computing}

\textbf{Qubits}: Qubits are the basic elements for quantum computing. In classic computing, a bit can only represent 0 or 1 at a time. However, a qubit can represent the superposition of 0 and 1. Formally $|\psi\rangle=\alpha|0>+\beta|1>$, where $\alpha, \beta \in \mathbb{C}$, $|\alpha|^{2}+|\beta|^{2}=1$.

\textbf{Quantum Measurement}: Quantum measurement is an approach for acquisition of a quantum state's information. The quantum measurement can collapse to $|0>$ or $|1>$.

\textbf{Quantum logic gate}: A quantum gate can be seen as a unitary transformation to qubits. The quantum logic gate should be a revertible gate. To support universal quantum computing, we only need to implement several single qubit's unitary transformation and a double gate (CNOT). The single qubit gates include the Hadamard gate, T gate and S gate. Widely used quantum logic gates and their corresponding unitary matrix are shown in Table. \ref{gatesMatrix}. 

\begin{table*}
	\begin{center}
		\begin{tabular}{|l|l|}
		\hline	 
		\makecell[c]{\textbf{Single qubit logic gate}} 
		&\makecell[c]{\textbf{Multi qubit logic gate}}  \\
		\hline
		& \\
       \makecell[c]{ H = $\begin{pmatrix}
	    	1/\sqrt{2} & 1/\sqrt{2} \\
	    	1/\sqrt{2} & -1/\sqrt{2} 
	    \end{pmatrix}$  }  
	    & 
	    \makecell[c]{CNOT = $\begin{pmatrix}
	    	1 & 0 & 0 & 0\\
	    	0 & 1 & 0 & 0\\
	    	0 & 0 & 0 & 1\\
	    	0 & 0 & 1 & 0\\
	    \end{pmatrix}$ }   \\
    		& \\
    		\hline 
    		& \\
	   \makecell[c]{ S = $\begin{pmatrix}
	    	1 & 0 \\
	    	0 & i \\
	    \end{pmatrix}$ } &\makecell[c]{ CZ = $\begin{pmatrix}
	      	1 & 0 & 0 & 0\\
	      	0 & 1 & 0 & 0\\
	      	0 & 0 & 1 & 0\\
	      	0 & 0 & 0 & -1\\
	      \end{pmatrix}$  }    \\ 
      		& \\ 
      		\hline
      		& \\
	   \makecell[c]{ T = $\begin{pmatrix}
	    	1 & 0 \\
	    	0 & e^{i\pi/4} \\
	    \end{pmatrix}$ } & \makecell[c]{ SWAP = $\begin{pmatrix}
	      	1 & 0 & 0 & 0\\
	      	0 & 0 & 1 & 0\\
	      	0 & 1 & 0 & 0\\
	      	0 & 0 & 0 & 1\\
	      \end{pmatrix}$ }  \\
      		& \\
      		\hline
      		& \\    
	  \makecell[c]{  X = $\begin{pmatrix}
	    	0 & 1 \\
	    	1 & 0 \\
	    \end{pmatrix}$}
	      & \makecell[c]{ CU = $\begin{pmatrix}
	      	1 & 0 & 0 & 0\\
	      	0 & 1 & 0 & 0\\
	      	0 & 0 & U_{00} & U_{01}\\
	      	0 & 0 & U_{10} & U_{11}\\
	      \end{pmatrix}$  }   \\  
	      		& \\  
	      		\hline
	      		& \\
	  \makecell[c]{  Y = $\begin{pmatrix}
	    	0 & -i \\
	    	i & 0 \\
	    \end{pmatrix}$ } & \makecell[c]{ Toffoli = $\begin{pmatrix}
	      	1 & 0 & 0 & 0 & 0 & 0 & 0 & 0\\
	      	0 & 1 & 0 & 0 & 0 & 0 & 0 & 0\\
	      	0 & 0 & 1 & 0 & 0 & 0 & 0 & 0\\
	      	0 & 0 & 0 & 1 & 0 & 0 & 0 & 0\\
	      	0 & 0 & 0 & 0 & 1 & 0 & 0 & 0\\
	      	0 & 0 & 0 & 0 & 0 & 1 & 0 & 0\\
	      	0 & 0 & 0 & 0 & 0 & 0 & 0 & 1\\
	      	0 & 0 & 0 & 0 & 0 & 0 & 1 & 0\\
	      \end{pmatrix}$  }    \\
	      	      		& \\  
	      \hline
	      & \\
	  \makecell[c]{ Z = $\begin{pmatrix}
	   	1 & 0 \\
	   	0 & -1 \\
	   \end{pmatrix}$ }  &\makecell[c]{ CR = $\begin{pmatrix}
	      	1 & 0 & 0 & 0\\
	      	0 & 1 & 0 & 0\\
	      	0 & 0 & 1 & 0\\
	      	0 & 0 & 0 & e^{i\theta}\\
	      \end{pmatrix}$   }   \\
            		& \\    
	    \hline     
         & \\
	  \makecell[c]{ $U_3$ = $\begin{pmatrix}
	   	cos(\frac{\theta}{2}) & -e^{i\lambda} \times sin(\frac{\theta}{2}) \\
	   	e^{i\phi} \times sin(\frac{\theta}{2}) & e^{i\lambda+i\phi} \times cos(\frac{\theta}{2}) \\
	   \end{pmatrix}$ }  &\makecell[c]{ iSWAP = $\begin{pmatrix}
	      	1 & 0 & 0 & 0\\
	      	0 & cos(\frac{\theta}{2}) & i sin(\frac{\theta}{2}) & 0\\
	      	0 & i sin(\frac{\theta}{2}) & cos(\frac{\theta}{2}) & 0\\
	      	0 & 0 & 0 & 1\\
	      \end{pmatrix}$   }   \\
            		& \\    
	    \hline       
		\end{tabular}
		\label{gatesMatrix}

	\end{center}
	
\end{table*}

\begin{figure}[htbp]
   \subfigure{\includegraphics[width=0.45\textwidth]{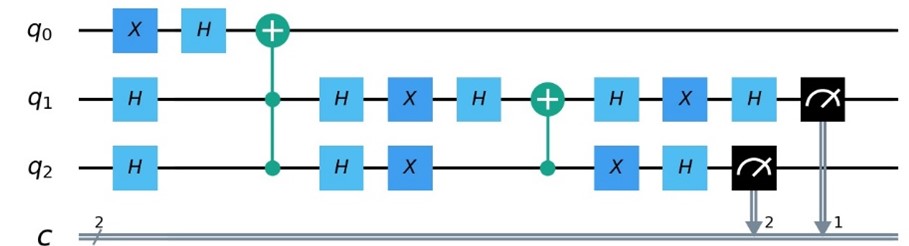}}
   \caption{The quantum circuit for the Grover algorithm}
   \label{GroverCircuit}
 \end{figure}

\textbf{Quantum circuit}: Quantum circuit is one of the most widely used quantum computing model. In this model, any unitary transformation can be implemented by combining several universal quantum gates. A sequence of quantum gates is called a ``quantum circuit". A quantum circuit can be visualized. For instance, the quantum circuit for the Grover algorithm can be represented as Fig. \ref{GroverCircuit}.

Generally, the initial quantum states are all zeroed in most quantum computing cases. After execution of a quantum circuit, we can get the results by measuring the qubits.

\textbf{Quantum Program}: A quantum program is consisted of a combination of quantum logic gates, classic computing and measurement.

\textbf{Quantum state's fidelity}: Due to the interference of noise and the quality of a quantum processor, the ideal results of quantum gate is not exactly the same as the real execution of quantum gates. The difference between the ideal quantum state and real quantum state can be represented as ``fidelity". The higher the fidelity means the less error, which also means the computing results tend to be better.

\subsection{Difference between classic computing and quantum computing}

The quantum computing and classic computing are based on different phyiscal theories. Thus, their paradigm and architecture are different. In this section, we discuss the main difference between quantum computing and classic computing to clarify why existing classic operating system cannot be directly applied to quantum computing. 

\subsubsection{Qubit Mapping}  

The topology and basic logic gates of different quantum processors are different. Thus, the different qubit mapping strategies can make the fidelity of the results given by a certain quantum circuit different during compilation of a quantum program. Moreover, the qubits' properties can also vary with time, which can make the qubit mapping more difficult. 

\subsubsection{Multi-Processor Scheduling}   

In a classic computer, the cores in a CPU are mostly homogeneous. Thus, when scheduling tasks on these cores, we don't need to consider the difference between instructions. Different physical implementations of a quantum computer can have different topology and support different quantum gates. After compilation, even the same circuit can be compiled as different quantum circuits under different topology and quantum gates. Thus, we should carefully schedule the quantum tasks in the quantum processors to better enable the performance.

\subsubsection{Quantum Parallel Computing}

Thread is the basic scheduling unit in a classic operating system. A single core can only execute a thread at a time. By switching between contexts, multiple threads can be easily switched with each other. Thus, multiple threads parallelism can be realized.

In a quantum computer, the quantum circuit is the basic scheduling unit. Since the state of a qubit cannot be cloned and the decoherence time of qubits is short, multiple quantum circuits cannot switch with each other as classic computing. However, when the qubits used by quantum circuits are different, the execution of multiple circuits can be realized in a quantum processor.

\subsubsection{Automatic Calibration}

The manufacture of physical instruments in classic computing is quite mature. The quality of these devices is stable. Their performance will not fluctuate in a short period of time. The objective of classic operating system is to improve the resource utilization through technologies such as memory management. For a quantum computer, we can improve the resource utilization through the parallel execution of quantum circuits. Existing solutions are based on an assumption that the quality of qubits are stable for a period of time. However, the quality of qubits will deteriorate during the execution of quantum circuits. In such situations, the qubit mapping cannot achieve satisfiable results with static single gate's fidelity, double gate's fidelity and measurement fidelity. When using quantum computers we should continuously check the quality of qubits and calibrate the qubits automatically.

In summary, the classic computing and quantum computing are quite different. Thus, classic operating systems cannot be easily compatible with a quantum computer. To meet this ends, Origin Pilot provides quantum task scheduling, quantum resource management, qubits' calibration, quantum circuits compiling to overcome the problem of qubits' automatic calibration and multiple quantum processors' load balancing. With Origin Pilot, the quantum circuits' fidelity and resource utilization of quantum resources can be greatly improved.

\subsection{Basic Definitions}

\textbf{Quantum application}: A quantum application is a hybrid program including both the classic computing part and quantum computing part;

\textbf{Quantum Task}: A quantum application can send multiple quantum circuits to quantum processors. Each quantum circuit can be seen as an individual quantum task. Thus, we abstract a quantum circuit as a quantum task.

\textbf{Quantum Transaction} A quantum transaction is the basic element which can be executed on a single quantum processor. It can includes multiple quantum tasks. These quantum tasks in a quantum transaction can be executed wholy or not. 

\textbf{Quantum Thread}: The basic unit for scheduling in a quantum operating system. The quantum operating system schedules the quantum threads based on the resource requirement of a quantum transaction. Once a quantum transaction is executed on a quantum processor, we can call it a ``quantum thread''.

\textbf{Quantum processor}: The basic execution unit of a quantum computer. A quantum processor can only execute a quantum thread at a time. A quantum application can use multiple quantum processors.

\textbf{Quantum programming framework}: A fundamental framework for building, excuting and optimizing a quanutm application. The framework can also provide basic algorithmatic libraries.

\textbf{Quantum resource}: Quantum resources refer to the physical system for processing and storing quantum information, following the rule of quantum mechanics. Specifically, the quantum resource includes the quantum processor and quantum storage. 

\section{Architecture of Origin Pilot} 

\subsection{Overall Architecture}

\begin{figure*}[htbp]
   \subfigure{\includegraphics[width=0.85\textwidth]{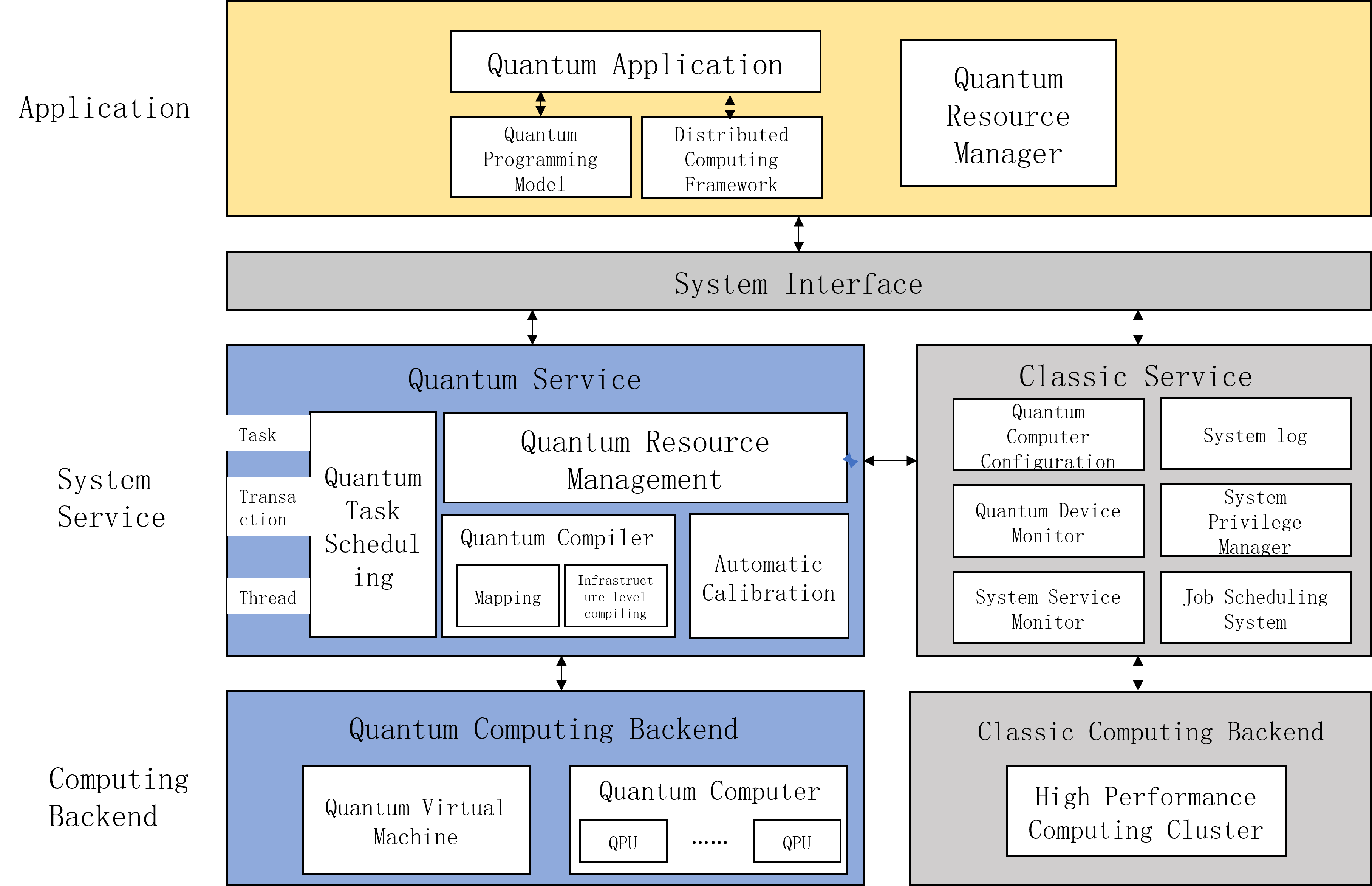}}
   \caption{Overall Arhictecture of Origin Pilot}
   \label{Origin PilotArch}
 \end{figure*}

Origin Pilot can support different computing backends such as quantum processors, quantum virtual machines and high performance computing clusters etc. The quantum computing needs the assistance of classic computers. For instance, when solving NP-hard problems, we should use classic computers to validate the results. For hybrid algorithms like quantum machine learning, quantum chemistry and quantum finance algorithms, the classic computing part plays a vital role. Thus, we should deal with classic information during the meanwhile execution of quantum tasks. Thus, we classify the system services to quantum services and classic services, which are shown in Fig \ref{Origin PilotArch}.

The quantum services are repsonsible for dealing with quantum tasks and interact with the quantum computing backend. By supporting multiple quantum processors' task scheduling, quantum resource management, multiple quantum circuit parallism, quantum program compilation and automatic calibration of qubits, we can improve the resource utilization of quantum processors and keep the fidelity of qubits within a certain threshold.

The classic computing services are dealing with classic computing tasks and interact with the classic computing backend. Moreover, the classic service should also be responsible for quantum computer's configuration, quantum device monitoring and system service monitoring. Specifically, large scale data processing with quantum and classic hybrid algorithms can be enabled with these classic services. The classic services should also monitor the state of quantum devices and system service to maintain the stability of the system.

A quantum application can call the system services provided above. Based on the quantum programming framework and distributed computing framework, Origin Pilot can support the quantum and classic hybrid distributed computing. A quantum application can preprocess the data with classic services. The quantum programs can be generated and sent to quantum processors. After computation, results can be retrieved by measurements and analyzed by classic computers. Further, we can determine the next parameterized quantum task. The users can also manage quantum devices and quantum resources with the resource manager in Origin Pilot.

\subsection{Workflow of Origin Pilot}

\begin{figure*}[htbp]
   \subfigure{\includegraphics[width=0.85\textwidth]{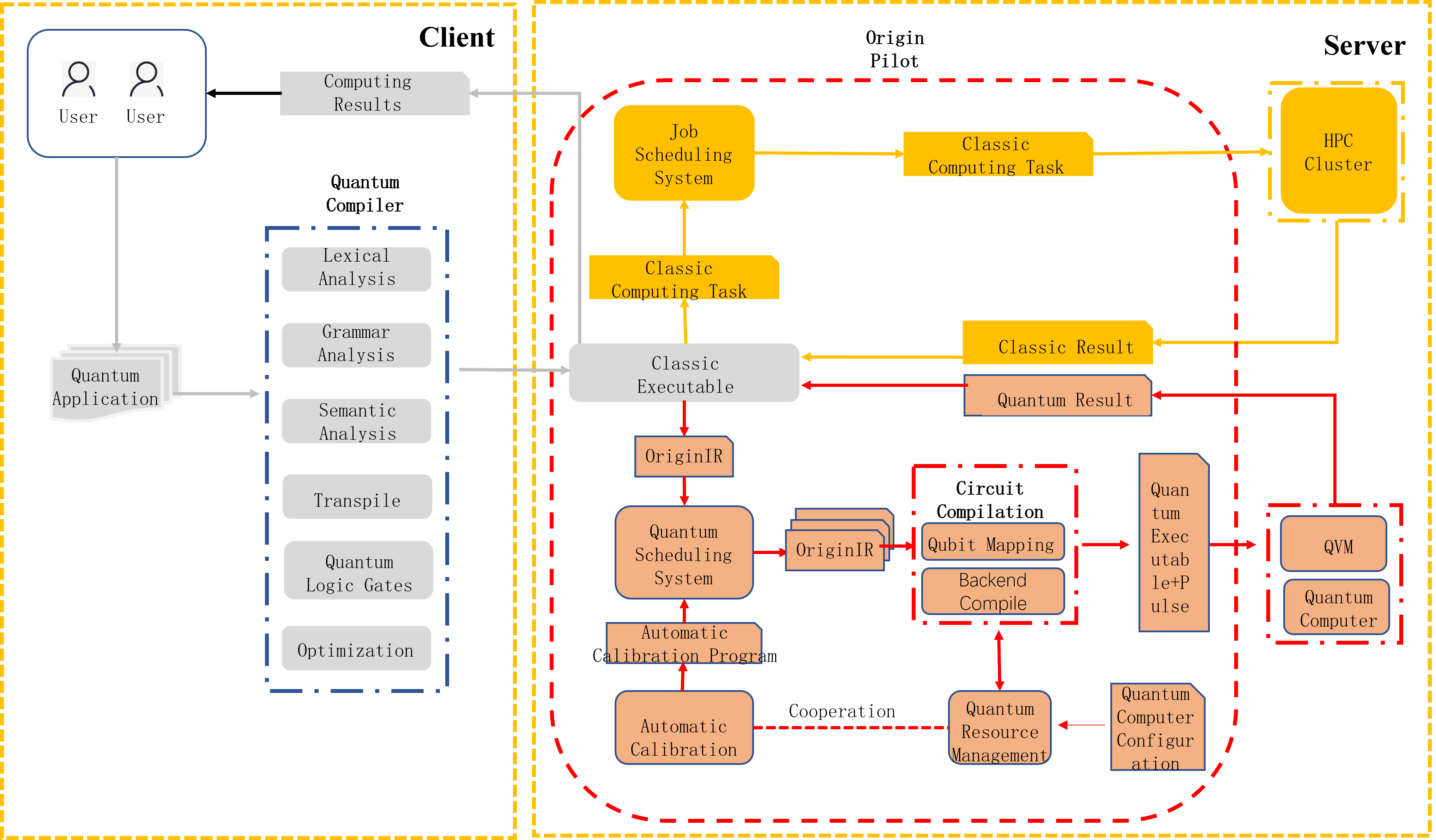}}
   \caption{Workflow of Origin Pilot}
   \label{PilotWorkflow}
 \end{figure*}

The workflow of Origin Pilot is shown in Fig.\ref{PilotWorkflow}. 

1. Users can write hyrbid programs with QRunes\cite{2019QRunes}. The QRunes compiler can identify the quantum part and classic part with lexical, grammer and semantic analysis. Then the hybrid quantum program can be transpiled to a quantum application which can be executed on the server side of Origin Pilot. If a hybrid program wants to use the high performance computing clusters, users can also program with distributed computing frameworks for the classic part. 

2. When Origin Pilot receives a hybrid quantum application, the classic part can be executed on the controlling server. If the hybrid program is written in distributed computing programming languages, the Origin Pilot will send the classic computing tasks to high performance computing clusters with a classic job scheduling system.

3. The quantum computing part will be sent to the quantum task scheduling service. The quantum task scheduling service will sort the quantum tasks based on their priority and choose a quantum task with the highest priority when its resource requirement can be met. Then the quantum circuit will be compiled to the topology of the target quantum processor.  Then the compiled code will be sent to the quantum computer. Before execution, a quantum transaction will bind to a quantum thread which will record the quantum transaction's ID, the target quantum processor's ID, the tasks's ID etc. After computing, we can identify the corresponding result for a quantum task and return it to the users' program. Then the occupied qubits will be released.

When executing the quantum tasks, Origin Pilot can also calibrate the performance of quantum resources. When the performance of qubits deteriorate, the automatic calibration service will set the qubits to be under-calibrated and notify the quantum task scheduling service to calibrate their state. Origin Pilot will assign highest priority to the calibration task and combine them with other quantum computing tasks as a quantum transaction which will be sent to the quantum devices.

\section{Solutions of Origin Pilot}

In this section, we will describe the solutions to the problem of multiple quantum processors' load balancing, multi-quantum program parallism and automatic calibration.

\subsection{Load Balancing of Multiple Quantum Processors}

The multiple quantum processors' load balancing is to schedule multiple quantum tasks on multiple quantum processors. In Origin Pilot, a quantum task is consisted of the following elements:

(1) The number of qubits required for a quantum task;

(2) Quantum program's intermediate representation;

(3) Quantum processor's ID;

(4) Type of a quantum task: a quantum task can be a general quantum task or automatic calibration task;

(5) Priority: The priority of a quantum task.

A quantum task can only be executed on corresponding the quantum processor if the quantum processor's ID is assigned. Otherwise, the quantum task scheduling system will allocate the quantum processors based on the system's state. Different scheduling algorithms will be applied based on the type of a quantum task.

A qubits' automatic calibration task usually requires real time response. The runtime of these tasks are usually very short. The physical qubits to be calibrated are explicitly described in a quantum task. To guarantee the reliability of the quantum computing, these type of quantum tasks should be prior to be executed. For general quantum tasks, we can allocate qubit resources based on the topology of quantum processors and system's state.We apply the HRRN (Highest Response Ratio Next) scheduling algorithm for the general quantum tasks. The HRRN algorithm considers both the waiting time and runtime of quantum tasks. The priority of a quantum task is defined as:

\begin{equation}
\mathrm{R}_{\mathrm{p}}=\frac{T_{ waiting\_time }+T_{ runtime }}{T_{ runtime }}=
\frac{T_{ response\_time }}{T_{ runtime }}
\end{equation}

With the increase of waiting time $T_{ waiting\_time }$, the quantum task with higher $R_p$ is prior to be executed. 

\begin{figure}[htbp]
   \subfigure{\includegraphics[width=0.45\textwidth]{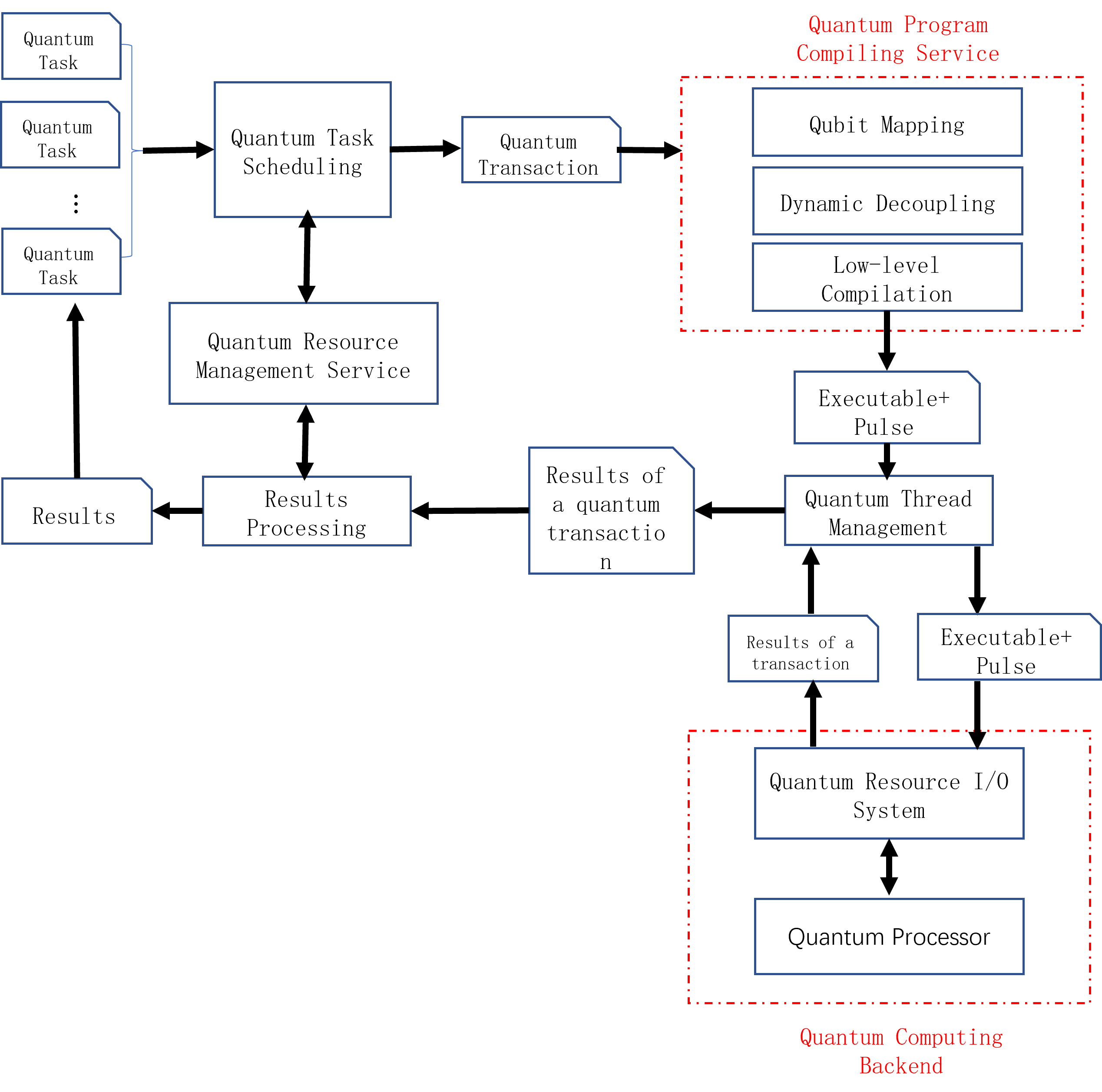}}
   \caption{Multiple quantum processors' load balancing}
   \label{loadBalance}
 \end{figure}

Based on the above algorithm, the workflow of quantum task scheduling service is shown in Fig.\ref{loadBalance}.   

Once receiving a quantum task, the quantum task scheduling service will:

(1) acquire the runtime of the quantum task and put the task into the waiting list;

(2) update the $R_p$ of all the tasks in waiting list;

(3) allocate quantum processors that can best fit the quantum task;

(4) compile the quantum task to the quantum executable file and pulse;

(5) bind the quantum task to a quantum thread and start execution.

Specifically, the qubits' automatic calibration task are assigned with higher priority. The qubits used are explicitly designated so the qubit mapping will not happen in this type of quantum tasks. Since the qubits in these quantum processors are divided into different regions, the automatic calibration tasks won't interfere the execution of general quantum tasks. Moreover, the calibration quantum tasks are scheduled with a FCFS(First-Come-First-Served) strategy. 

\subsection{Parallel Execution of Multiple Quantum Programs}

Currently Origin Pilot apply synchronous parallisation to enable parallel computing. Multiple quantum tasks are combined as a quantum transaction. The quantum tasks in the quantum transaction will be executed in parallel. Each time only one quantum transaction can be executed. Before execution, the logic qubits will be mapped to the physical qubits in the quanutm processors. 

The problem of mapping is to transform a quantum circuit to a target quantum circuit that can be directly executed on a quantum processor while maintaining its original function\cite{2019Tackling}\cite{2019Noise}\cite{2019Mapping}. Sometimes a quantum circuit may require two qubits to be entangled. However, in a physical quantum processor, the qubits cannot directly communicate with each other. To tackle the problem, we introduce a series of SWAP gates. The SWAP gates can be decomposed to basic gates supported by a quantum processor. Although the theoretical results are the same, extra error may be incurred due to the extra SWAP gates. Thus, we try to minimize the number of SWAP gates when mapping user's quantum circuits to physical quantum processors.

The qubits' mapping can be seen as a token-swap problem\cite{E2016Complexity}. As the number of qubits increases, the time complexity for finding the optimal mapping solution increases exponentially. Apart from minimizing the SWAP gates, we should also consider the difference between qubits. Thus, we should choose the route with the best fidelity. 

\begin{algorithm}[H]
\begin{algorithmic}[1]
\Require src\_QProg (Original quantum program), QPU\_adj (topology of a quantum processor)
\Ensure mapped\_QC (a quantum program that can fit in the input quantum processor after mapping)
\State Convert the original quantum program to a DAG;
\State Initialize the sub\_graph\_vec\_2d to store the maximum subgraph sequence;\\
\Comment{phase\_1: Traverse the DAG to get the maximum subgraph sequence}
\While{(the vertex number of the DAG$>$0)}
\State Choose the vertex V with in-degree=0;
\If{(the sequence of subgraph S is not NULL)}
\If{(all the subgraph of S is not extensible)}
\State Append S to sub\_graph\_vec\_2d;
\State Clear the elements in S;
\State Break;
\Else
\State Extend S based on QPU\_adj; \par
\hskip\algorithmicindent Remove the un-extensible subgraph from S;
\EndIf
\Else
\State Initialize S based on the possible mapping from V to QPU\_adj;
\EndIf
\EndWhile\par
\Comment{phase\_2:Token-Swapping, get the best path}
\State Initialize best\_path\_vec to store the best path;
\ForEach{S\_cur in sub\_graph\_vec\_2d}
\State Calculate the minimum Cost from each subgraph of S\_cur to each subgraph of S\_next with Token-Swapping algorithm;
\State Append the best path $best\_swap$ to $sub\_best\_path\_vec$;
\EndFor
\ForEach{$best\_path \text{ in } best\_path\_vec$}
\State Calculate the overall fidelity of the $best\_path$
\EndFor
\State Choose the $final\_best_path$ with best fidelity from $best\_path\_vec$;\par
\Comment{phase\_3:Traverse $final\_best\_path$, and generate the new mapped\_QC}
\ForEach{$path\_node in final\_best\_path$}
\If{($path\_node$ is a subgraph)}
\State Convert $path\_node$ to a quantum program $sub\_cir$;
\State Insert $sub\_ cir$ to $mapped\_QC$;
\EndIf
\If{(path\_node is best\_swap)}
\State Insert $swap-gates$ to $mapped\_QC$;
\EndIf
\EndFor\par
\Return mapped\_QC;
\end{algorithmic}
\end{algorithm}

Pseudo code of the mapping algorithm is shown above. (1) We first convert the quantum program to a DAG(Directed Acyclic Graph). The vertices represent the double gate operation. When two vertices are connected, the corresponding double gates will use the same qubit. The direction denotes the timing sequence. (2) We traverse the DAG from the node with 0 in-degree. The first node will be directly mapped based on the topology of the quantum processor. Each mapping will be represented as a subgraph. (3) We get the new DAG by deleting the mapped vertices. Then we traverse the new DAG until all the nodes with 0 in-degree cannot be directly mapped. In this way, we can finally get the maximum subgraph sequence. (4) By repeating step (2) and (3), we can get multiple maximum subgraph sequences. (5) With Token-Swapping algorithm, we can calculate the path with the least SWAP gates of multiple mapping strategies by connecting the adjacent subgraphs. (6) We elaborate all the possible mapping strategies and choose the mapping with best fidelity.

\subsection{Automatic Calibration}

There are two parameters describing the quality of qubits: coherence time and gate fidelity.

Qubits' coherence time can be used to dscribe the coupling strength between a quantum system and the environment. A quantum algorithm may need a massive number of gate operations. Thus, the qubits should maintain their state during these gate operations. 

Quantum logic gate operations serve as the basic elements in a quantum circuit. The gate error has a great impact on the final result of a quantum circuit. Generally, the average logic gate error should be less than 1 percent\cite{2015State}. 

There are many factors that can affect the qubits' quality. To maintain the quality, we need to keep calibrating the qubits. The qubits' calibration includes the checking and calibration phase. In the checking phase, we can check the qubits' performance parameters by interval checking or random traverse techniques. If calibration is needed, we can call the corresponding calibration procedure based on the error type and extent. 

The factors that can affect the qubits' performance are co-related with each other. To conduct an effective calibration, we need to determine which phyiscal parameters degrade the qubits' performance. This process can be formulated as a Markov decision process and automated. We build a partial observable Markov Decision Process to automate the process of automatic calibration.

Moreover, we build an oline calibration strategy by using a block partitioning automatic calibration technique. The approach dynamically divide the quantum processor to the executable region and calibration region. The quantum program can be compiled to different regions and combined as a quantum transaction which can be further sent to the quantum processors. With the above framework, we can assure the user-submitted quantum tasks being executed on the best region of the quantum processor. Thus, the fidelity of the result of the quantum task tends to be better.

\section{Experiments}

We conduct several exepriments to evaluate the different aspects of Origin Pilot's effectiveness.

\subsection{Runtime Analysis}

To evaluate the effectiveness of the parallesm in Origin Pilot, we conduct the runtime analysis with and without Origin Pilot on two superconducting quantum processors (KF C6-130) provided by Origin Quantum. 

\begin{figure}[htbp]
   \subfigure{\includegraphics[width=0.45\textwidth]{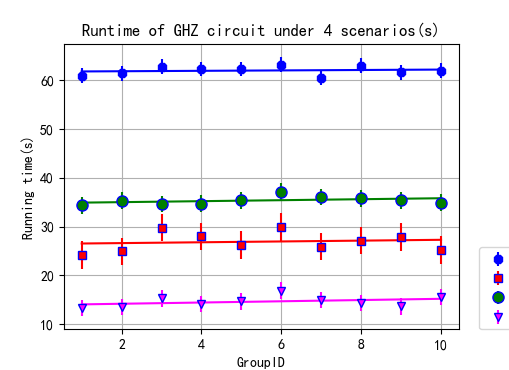}}
   \caption{Runtime of four different situations}
   \label{runtime}
 \end{figure}

We conduct four exepriments in total. In the first scenario, we execute a single quantum circuit of GHZ in a quantum chip for 10 times. The GHZ circuit uses 2 qubits. In the second scenario, we execute two quantum circuits in a quantum chip. In the third scenario, we run a quantum circuit in two quantum chips, taking advantage of the parallelism of Origin Pilot. In the fourth scenario, we execute the two quantum circuits in two quantum chips. From Fig.\ref{runtime} we can see the acceleration of Origin Pilot by running the above four scenarios for 10 times. 

Origin Pilot can effectively improve the utilisation of quantum processors by supporting parallel execution of multiple tasks in a quantum processor. The average runtime of executing a GHZ quantum circuit can be improved to 120 percent.

\subsection{Automatic Calibration Results}

All the results of this part are evaluated on the OriginQ Wuyuan. We mainly evaluate the single gate's and double gate's fidelity results with automatic calibration. Moreover, since the automatic calibration quantum tasks are also considered as quantum tasks executed on the quantum processors. Thus, we record the number of quantum tasks as well.

\subsubsection{Single gate results}

The calibration threhold is set to 0.98. When Origin Pilot detects the fidelity of single gate operation of a qubit below 0.98, the calibration procedure is triggered. The calibration interval is initially set as 60 minutes and gradually degraded to 20 minutes. Once the calibration is triggered, the interval is recovered to 60 minutes.

\begin{figure*}[htbp]
   \subfigure{\includegraphics[width=0.45\textwidth]{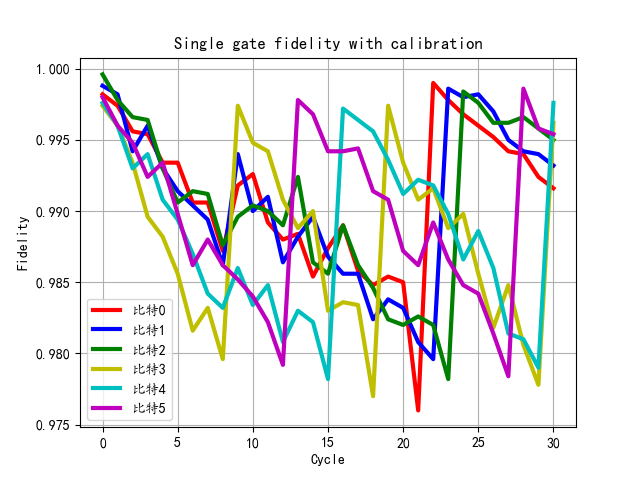}}
	\subfigure{\includegraphics[width=0.45\textwidth]{single-qubit-with-calibrate.png}}
   \caption{Fidelity Results of Single Qubit Calibration}
   \label{SingleQubitCalibrateResult}
 \end{figure*}

Experimental results are shown in Fig.\ref{SingleQubitCalibrateResult}.

\subsubsection{Double gate results}

Every 20 minutes we conduct our automatic calibration process and summarize the fidelity. We set the calibration threshold as 0.95. Initially the calibration is conducted every 60 minutes. The calibration interval is reduced to 20 minutes gradually. Once the calibration is conducted, the interval is adjusted back to 60 minutes. 

\begin{figure*}[htbp]
   \subfigure{\includegraphics[width=0.45\textwidth]{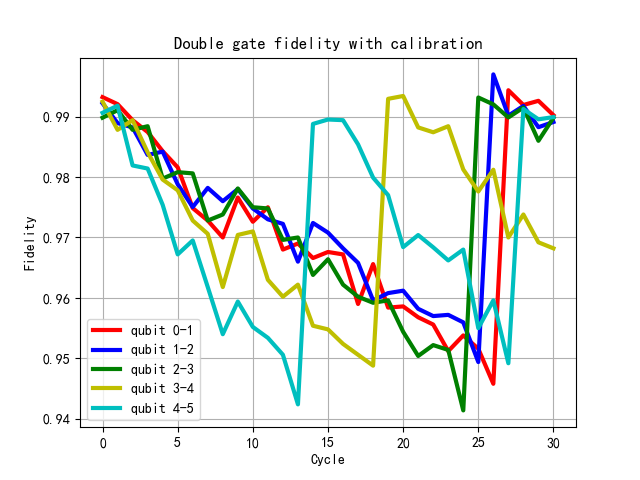}}
	\subfigure{\includegraphics[width=0.45\textwidth]{double-qubit-with-calibrate.png}}
   \caption{Fidelity Results of Double Qubit Calibration}
   \label{DoubleQubitCalibrateResult}
 \end{figure*}

Experimental results are shown in Fig.\ref{DoubleQubitCalibrateResult}

\subsubsection{Number of tasks comparison}

We generate a random GHZ quantum circuit. The quantum task interval is set to 10 seconds. We summarize the number of tasks in each period. The double gate fidelity threshold is set to 0.95; single gate fidelity threshold is set to 0.98. Exprimental results are shown as below in Fig. \ref{taskNum}.

 \begin{figure*}[htbp]
   \subfigure{\includegraphics[width=0.45\textwidth]{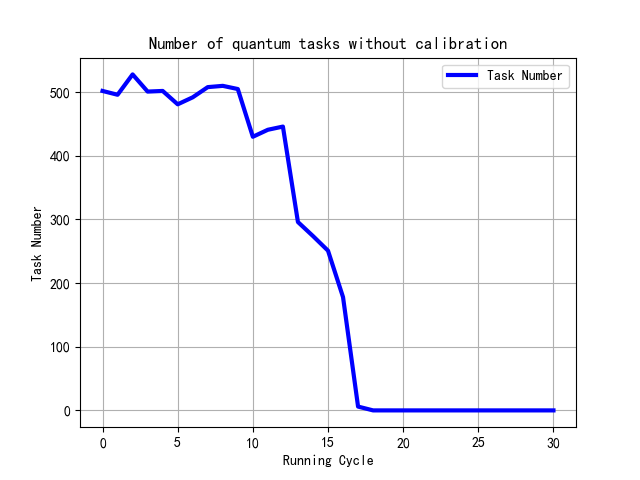}}
	\subfigure{\includegraphics[width=0.45\textwidth]{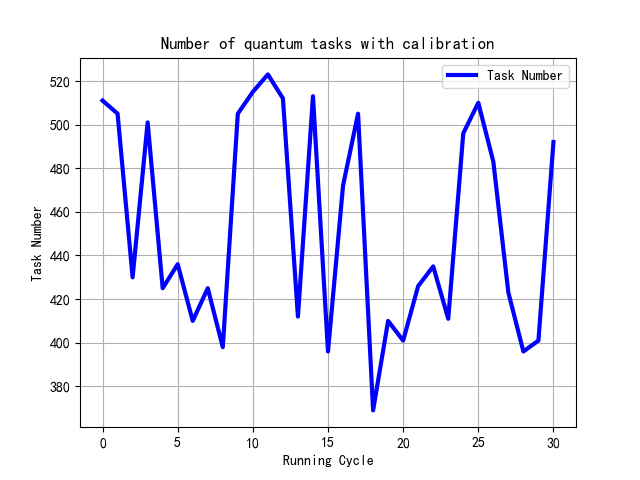}}
   \caption{Number of Tasks with and without Origin Pilot}
   \label{taskNum}
 \end{figure*}

\subsubsection{Results Analysis}

From the results we can see that without automatic calibration service, the single and double gate's fidelity degrade gradually. The automatic calibration service can keep the single and double gate's fidelity above 0.98 and 0.95 respectively. With automatic calibration, we can see the number of tasks is far more than the quantum tasks without calibration. Moreover, with automatic calibration service, the quantum processor can keep working properly. But the quantum processor without  automatic calibration can no longer support the execution of quantum tasks.

\subsection{Effect of the Qubit Mapping Mechanism in Origin Pilot}

\subsubsection{Dataset}

We build a topology of a quantum processor including 8 qubits. Double gate operations can be applied in adjacent qubits. We configure a quantum virtual machine with noise and only consider the CZ gate. The noise can be assigned to the corresponding qubits and CZ gate. To better validate our algorithm, we set the fidelity on the right part of the topology higher than the left. We use the classic quantum algorithms including QFT, GHZ, DJ and BV as the benchmarking quantum circuits. 

\begin{figure}[htbp]
  \subfigure{\includegraphics[width=0.45\textwidth]{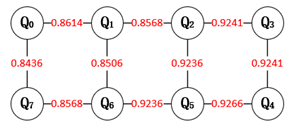}}
  \caption{Topology of a simulated quantum processor}
  \label{qubitsTopo}
\end{figure}

The topology of the quantum processor is shown as Fig.\ref{qubitsTopo}.

\subsubsection{Experimental Results}

(1) QFT Circuit

\begin{figure}[H]
 	\centering
 	\includegraphics[width=0.45\textwidth]{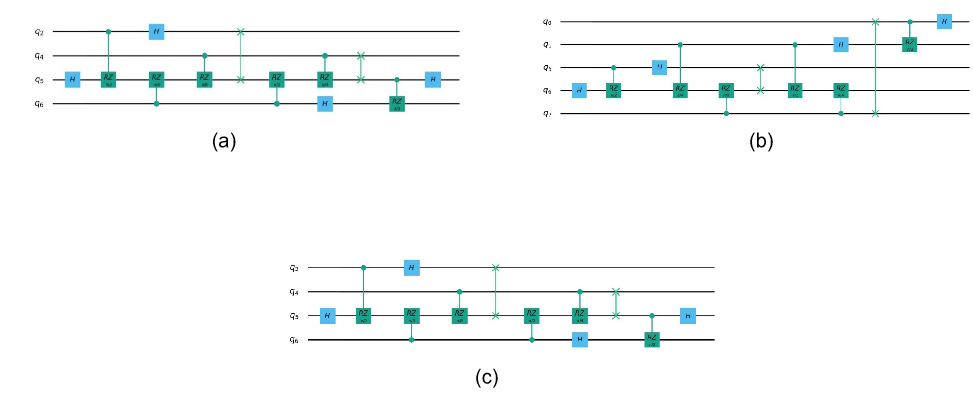}
 	\caption{Circuit of QFT}
	\label{QFTCircuit}
 \end{figure}

Fig. \ref{QFTCircuit}(a) shows the original circuit of QFT. Fig.\ref{QFTCircuit}(b) shows a transpiled circuit with BMT. Fig.\ref{QFTCircuit}(c) shows a transpiled circuit with Origin Pilot.

\begin{table}[H]
	\begin{center}
		\begin{tabular}{|c|c|c|c|c|c|}
			\hline  		 
			 & 1& 2& 3& 4&average \\
			\hline  
			Origin Pilot   &0.1261 &0.1307&0.1241&0.1253& 0.1266 \\
			\hline  
			BMT &0.2125&0.2152&0.2102&0.2178&0.2139 \\
			\hline  
		\end{tabular}
		\caption{Fidelity Results for QFT with and without Origin Pilot}
	\end{center}
\end{table}

(2) GHZ Circuit

\begin{figure}[H]
	\centering
	\includegraphics[width=0.45\textwidth]{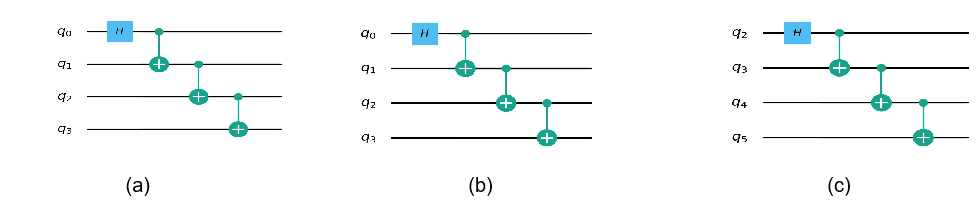}
	\caption{Circuit of GHZ}
	\label{GHZCircuit}
\end{figure}

Fig. \ref{GHZCircuit}(a) shows the original circuit of GHZ. Fig.\ref{GHZCircuit}(b) shows a transpiled circuit with BMT. Fig.\ref{GHZCircuit}(c) shows a transpiled circuit with Origin Pilot.

\begin{table}[H]
	\begin{center}
		\begin{tabular}{|c|c|c|c|c|c|}
			\hline  		 
			& 1& 2& 3& 4&average \\
			\hline  
			Origin Pilot   &0.435741 &0.42889&0.426278&0.431303& 0.450553 \\
			\hline  
			BMT &0.599643&0.604789&0.60461&0.604014&0.603264 \\
			\hline  
		\end{tabular}
		\caption{Fidelity Results for GHZ with and without Origin Pilot}
	\end{center}
\end{table}

(3) DJ Circuit

\begin{figure}[h]
	\centering
	\includegraphics[width=0.45\textwidth]{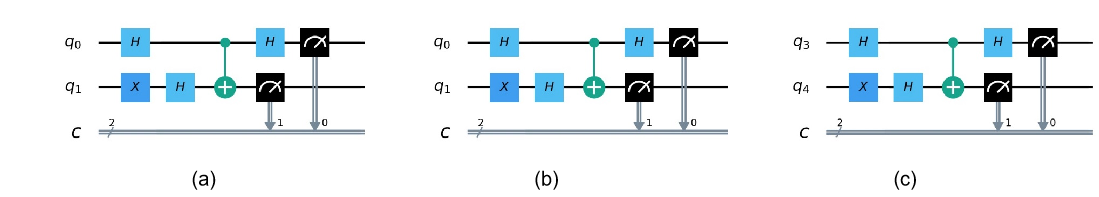}
	\caption{Circuit of DJ}
	\label{DJCircuit}
\end{figure}

Fig. \ref{DJCircuit}(a) shows the original circuit of DJ. Fig.\ref{DJCircuit}(b) shows a transpiled circuit with BMT. Fig.\ref{DJCircuit}(c) shows a transpiled circuit with Origin Pilot.

\begin{table}[H]
	\begin{center}
		\begin{tabular}{|c|c|c|c|c|c|}
			\hline  		 
			& 1& 2& 3& 4&average \\
			\hline  
			Origin Pilot   &0.435741 &0.42889&0.426278&0.431303& 0.450553 \\
			\hline  
			BMT &0.86&0.8536&0.8631&0.8619&0.85965 \\
			\hline  
		\end{tabular}
		\caption{Fidelity Results for DJ with and without Origin Pilot}
	\end{center}
\end{table}

(4) BV Circuit

\begin{figure}[H]
	\centering
	\includegraphics[width=0.45\textwidth]{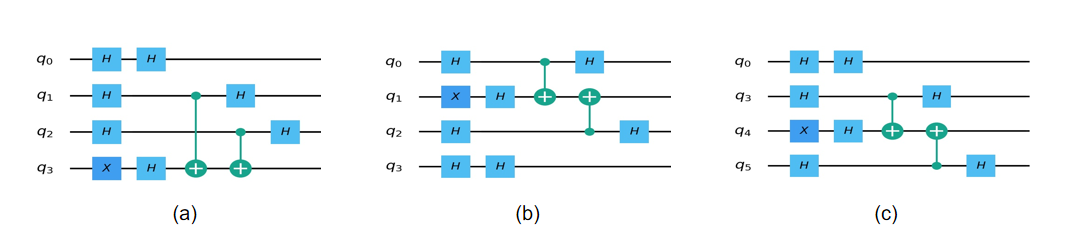}
	\caption{Circuit of BV}
	\label{BVCircuit}
\end{figure}

Fig. \ref{BVCircuit}(a) shows the original circuit of BV. Fig.\ref{BVCircuit}(b) shows a transpiled circuit with BMT. Fig.\ref{BVCircuit}(c) shows a transpiled circuit with Origin Pilot.

\begin{table}[H]
	\begin{center}
		\begin{tabular}{|c|c|c|c|c|c|}
			\hline  		 
			& 1& 2& 3& 4&average \\
			\hline  
			Origin Pilot  &0.547951 &0.542246&0.553503&0.5379& 0.5454 \\
			\hline  
			BMT &0.736669&0.729772&0.739268&0.743237&0.7372365 \\
			\hline  
		\end{tabular}
		\caption{Fidelity Results for BV with and without Origin Pilot}
	\end{center}
\end{table}

\subsubsection{Results' Analysis}

We can see that the fidelity from QST (Quantum State Tomography) shows that mapping with Origin Pilot outperforms mapping with BMT. In the worst case we can increase the fidelity of a quantum circuit by 10 percent on average. From the transpiled results we can also see that the mapping of Origin Pilot prefers choosing the qubits with better fidelity. Thus, the mapping with Origin Pilot can achieve better fidelity.

\section{Conclusion and Future Works}

The research on quantum operating systems is still in its infancy. Origin Pilot is a full quantum operating system among the primitive frameworks of quantum operating systems. We introduce in detail the implementation of basic modules of the quantum operating system, such as quantum task scheduling, quantum resource management,parallel execution and automatic calibration, etc.

In the future we will further support quantum distributed computing and hybrid computing consisting of both classic and quantum resources. We will also make the source code of Origin Pilot public available and make it free for the uncommecial usage of Origin Pilot.

\bibliography{ref}

\begin{thebibliography}{26}%
\makeatletter
\providecommand \@ifxundefined [1]{%
 \@ifx{#1\undefined}
}%
\providecommand \@ifnum [1]{%
 \ifnum #1\expandafter \@firstoftwo
 \else \expandafter \@secondoftwo
 \fi
}%
\providecommand \@ifx [1]{%
 \ifx #1\expandafter \@firstoftwo
 \else \expandafter \@secondoftwo
 \fi
}%
\providecommand \natexlab [1]{#1}%
\providecommand \enquote  [1]{``#1''}%
\providecommand \bibnamefont  [1]{#1}%
\providecommand \bibfnamefont [1]{#1}%
\providecommand \citenamefont [1]{#1}%
\providecommand \href@noop [0]{\@secondoftwo}%
\providecommand \href [0]{\begingroup \@sanitize@url \@href}%
\providecommand \@href[1]{\@@startlink{#1}\@@href}%
\providecommand \@@href[1]{\endgroup#1\@@endlink}%
\providecommand \@sanitize@url [0]{\catcode `\\12\catcode `\$12\catcode
  `\&12\catcode `\#12\catcode `\^12\catcode `\_12\catcode `\%12\relax}%
\providecommand \@@startlink[1]{}%
\providecommand \@@endlink[0]{}%
\providecommand \url  [0]{\begingroup\@sanitize@url \@url }%
\providecommand \@url [1]{\endgroup\@href {#1}{\urlprefix }}%
\providecommand \urlprefix  [0]{URL }%
\providecommand \Eprint [0]{\href }%
\providecommand \doibase [0]{http://dx.doi.org/}%
\providecommand \selectlanguage [0]{\@gobble}%
\providecommand \bibinfo  [0]{\@secondoftwo}%
\providecommand \bibfield  [0]{\@secondoftwo}%
\providecommand \translation [1]{[#1]}%
\providecommand \BibitemOpen [0]{}%
\providecommand \bibitemStop [0]{}%
\providecommand \bibitemNoStop [0]{.\EOS\space}%
\providecommand \EOS [0]{\spacefactor3000\relax}%
\providecommand \BibitemShut  [1]{\csname bibitem#1\endcsname}%
\let\auto@bib@innerbib\@empty
\bibitem [{\citenamefont {Nielsen}\ and\ \citenamefont
  {Chuang}(2002)}]{2002Quantum}%
  \BibitemOpen
  \bibfield  {author} {\bibinfo {author} {\bibfnamefont {M.~A.}\ \bibnamefont
  {Nielsen}}\ and\ \bibinfo {author} {\bibfnamefont {I.~L.}\ \bibnamefont
  {Chuang}},\ }\href@noop {} {\bibfield  {journal} {\bibinfo  {journal}
  {Mathematical Structures in Computer Science}\ }\textbf {\bibinfo {volume}
  {17}},\ \bibinfo {pages} {1115} (\bibinfo {year} {2002})}\BibitemShut
  {NoStop}%
\bibitem [{\citenamefont {Grover}(1996)}]{1996A}%
  \BibitemOpen
  \bibfield  {author} {\bibinfo {author} {\bibfnamefont {L.~K.}\ \bibnamefont
  {Grover}},\ }\href@noop {} {\  (\bibinfo {year} {1996})}\BibitemShut
  {NoStop}%
\bibitem [{\citenamefont {Harrow}\ \emph {et~al.}(2009)\citenamefont {Harrow},
  \citenamefont {Hassidim},\ and\ \citenamefont {Lloyd}}]{2009Quantum}%
  \BibitemOpen
  \bibfield  {author} {\bibinfo {author} {\bibfnamefont {A.~W.}\ \bibnamefont
  {Harrow}}, \bibinfo {author} {\bibfnamefont {A.}~\bibnamefont {Hassidim}}, \
  and\ \bibinfo {author} {\bibfnamefont {S.}~\bibnamefont {Lloyd}},\
  }\href@noop {} {\bibfield  {journal} {\bibinfo  {journal} {Physical Review
  Letters}\ }\textbf {\bibinfo {volume} {103}},\ \bibinfo {pages} {150502}
  (\bibinfo {year} {2009})}\BibitemShut {NoStop}%
\bibitem [{\citenamefont {Shor}(1994)}]{1994Algorithms}%
  \BibitemOpen
  \bibfield  {author} {\bibinfo {author} {\bibfnamefont {P.}~\bibnamefont
  {Shor}},\ }\href@noop {} {\bibfield  {journal} {\bibinfo  {journal} {In
  Proceedings of 35th Annual Symposium on the Foundations of Computer Science,
  IEEE Computer Society Press, Los Alamitos, CA}\ ,\ \bibinfo {pages} {124}}
  (\bibinfo {year} {1994})}\BibitemShut {NoStop}%
\bibitem [{\citenamefont {Peruzzo}\ \emph {et~al.}(2013)\citenamefont
  {Peruzzo}, \citenamefont {Mcclean}, \citenamefont {Shadbolt}, \citenamefont
  {Yung}, \citenamefont {Zhou}, \citenamefont {Love}, \citenamefont
  {Aspuru-Guzik},\ and\ \citenamefont {O'Brien}}]{2013A}%
  \BibitemOpen
  \bibfield  {author} {\bibinfo {author} {\bibfnamefont {A.}~\bibnamefont
  {Peruzzo}}, \bibinfo {author} {\bibfnamefont {J.}~\bibnamefont {Mcclean}},
  \bibinfo {author} {\bibfnamefont {P.}~\bibnamefont {Shadbolt}}, \bibinfo
  {author} {\bibfnamefont {M.~H.}\ \bibnamefont {Yung}}, \bibinfo {author}
  {\bibfnamefont {X.~Q.}\ \bibnamefont {Zhou}}, \bibinfo {author}
  {\bibfnamefont {P.~J.}\ \bibnamefont {Love}}, \bibinfo {author}
  {\bibfnamefont {A.}~\bibnamefont {Aspuru-Guzik}}, \ and\ \bibinfo {author}
  {\bibfnamefont {J.~L.}\ \bibnamefont {O'Brien}},\ }\href@noop {} {\bibfield
  {journal} {\bibinfo  {journal} {Nature Communications}\ }\textbf {\bibinfo
  {volume} {5}} (\bibinfo {year} {2013})}\BibitemShut {NoStop}%
\bibitem [{\citenamefont {Ghosh}\ and\ \citenamefont
  {Kozarevic}(2018)}]{2018Identifying}%
  \BibitemOpen
  \bibfield  {author} {\bibinfo {author} {\bibfnamefont {B.}~\bibnamefont
  {Ghosh}}\ and\ \bibinfo {author} {\bibfnamefont {E.}~\bibnamefont
  {Kozarevic}},\ }\href@noop {} {\bibfield  {journal} {\bibinfo  {journal}
  {Investment Management and Financial Innovations}\ }\textbf {\bibinfo
  {volume} {15}},\ \bibinfo {pages} {208} (\bibinfo {year} {2018})}\BibitemShut
  {NoStop}%
\bibitem [{\citenamefont {Biamonte}\ \emph {et~al.}()\citenamefont {Biamonte},
  \citenamefont {Wittek}, \citenamefont {Pancotti}, \citenamefont {Rebentrost},
  \citenamefont {Wiebe},\ and\ \citenamefont {Lloyd}}]{0Quantum}%
  \BibitemOpen
  \bibfield  {author} {\bibinfo {author} {\bibfnamefont {J.}~\bibnamefont
  {Biamonte}}, \bibinfo {author} {\bibfnamefont {P.}~\bibnamefont {Wittek}},
  \bibinfo {author} {\bibfnamefont {N.}~\bibnamefont {Pancotti}}, \bibinfo
  {author} {\bibfnamefont {P.}~\bibnamefont {Rebentrost}}, \bibinfo {author}
  {\bibfnamefont {N.}~\bibnamefont {Wiebe}}, \ and\ \bibinfo {author}
  {\bibfnamefont {S.}~\bibnamefont {Lloyd}},\ }\href@noop {} {\bibinfo
  {journal} {Nature}\ }\BibitemShut {NoStop}%
\bibitem [{\citenamefont {Petruccione}\ \emph {et~al.}(2015)\citenamefont
  {Petruccione}, \citenamefont {Francesco}, \citenamefont {Sinayskiy},
  \citenamefont {Ilya}, \citenamefont {Schuld},\ and\ \citenamefont
  {Maria}}]{Petruccione2015An}%
  \BibitemOpen
\bibfield  {journal} {  }\bibfield  {author} {\bibinfo {author} {\bibnamefont
  {Petruccione}}, \bibinfo {author} {\bibnamefont {Francesco}}, \bibinfo
  {author} {\bibnamefont {Sinayskiy}}, \bibinfo {author} {\bibnamefont {Ilya}},
  \bibinfo {author} {\bibnamefont {Schuld}}, \ and\ \bibinfo {author}
  {\bibnamefont {Maria}},\ }\href@noop {} {\bibfield  {journal} {\bibinfo
  {journal} {Contemporary Physics A Review of Physics and Associated
  Technologies}\ } (\bibinfo {year} {2015})}\BibitemShut {NoStop}%
\bibitem [{\citenamefont {Li}\ and\ \citenamefont {X.}(2003)}]{Li2003An}%
  \BibitemOpen
  \bibfield  {author} {\bibinfo {author} {\bibnamefont {Li}}\ and\ \bibinfo
  {author} {\bibnamefont {X.}},\ }\href@noop {} {\bibfield  {journal} {\bibinfo
   {journal} {Science}\ }\textbf {\bibinfo {volume} {301}},\ \bibinfo {pages}
  {809} (\bibinfo {year} {2003})}\BibitemShut {NoStop}%
\bibitem [{\citenamefont {Houck}\ \emph {et~al.}(2012)\citenamefont {Houck},
  \citenamefont {Türeci},\ and\ \citenamefont {Koch}}]{2012On}%
  \BibitemOpen
  \bibfield  {author} {\bibinfo {author} {\bibfnamefont {A.~A.}\ \bibnamefont
  {Houck}}, \bibinfo {author} {\bibfnamefont {H.}~\bibnamefont {Türeci}}, \
  and\ \bibinfo {author} {\bibfnamefont {J.}~\bibnamefont {Koch}},\ }\href@noop
  {} {\bibfield  {journal} {\bibinfo  {journal} {Nature Physics}\ }\textbf
  {\bibinfo {volume} {8}},\ \bibinfo {pages} {292} (\bibinfo {year}
  {2012})}\BibitemShut {NoStop}%
\bibitem [{\citenamefont {Haffner}\ \emph {et~al.}(2008)\citenamefont
  {Haffner}, \citenamefont {Roos},\ and\ \citenamefont {Blatt}}]{H2008Quantum}%
  \BibitemOpen
  \bibfield  {author} {\bibinfo {author} {\bibfnamefont {H.}~\bibnamefont
  {Haffner}}, \bibinfo {author} {\bibfnamefont {C.}~\bibnamefont {Roos}}, \
  and\ \bibinfo {author} {\bibfnamefont {R.}~\bibnamefont {Blatt}},\
  }\href@noop {} {\bibfield  {journal} {\bibinfo  {journal} {Physics Reports}\
  } (\bibinfo {year} {2008})}\BibitemShut {NoStop}%
\bibitem [{\citenamefont {Kielpinski}\ \emph {et~al.}(2002)\citenamefont
  {Kielpinski}, \citenamefont {Monroe},\ and\ \citenamefont
  {Wineland}}]{2002Architecture}%
  \BibitemOpen
  \bibfield  {author} {\bibinfo {author} {\bibnamefont {Kielpinski}}, \bibinfo
  {author} {\bibfnamefont {C.}~\bibnamefont {Monroe}}, \ and\ \bibinfo {author}
  {\bibfnamefont {D.~J.}\ \bibnamefont {Wineland}},\ }\href@noop {} {\bibfield
  {journal} {\bibinfo  {journal} {Nature}\ }\textbf {\bibinfo {volume} {417}},\
  \bibinfo {pages} {709} (\bibinfo {year} {2002})}\BibitemShut {NoStop}%
\bibitem [{\citenamefont {Lvovsky}\ \emph {et~al.}(2009)\citenamefont
  {Lvovsky}, \citenamefont {Sanders},\ and\ \citenamefont
  {Tittel}}]{2009Optical}%
  \BibitemOpen
  \bibfield  {author} {\bibinfo {author} {\bibfnamefont {A.~I.}\ \bibnamefont
  {Lvovsky}}, \bibinfo {author} {\bibfnamefont {B.~C.}\ \bibnamefont
  {Sanders}}, \ and\ \bibinfo {author} {\bibfnamefont {W.}~\bibnamefont
  {Tittel}},\ }\href@noop {} {\bibfield  {journal} {\bibinfo  {journal} {Nature
  Photonics}\ }\textbf {\bibinfo {volume} {3}},\ \bibinfo {pages} {706}
  (\bibinfo {year} {2009})}\BibitemShut {NoStop}%
\bibitem [{\citenamefont {O'Brien}(2007)}]{2007Optical}%
  \BibitemOpen
  \bibfield  {author} {\bibinfo {author} {\bibfnamefont {J.~L.}\ \bibnamefont
  {O'Brien}},\ }\href@noop {} {\  (\bibinfo {year} {2007})}\BibitemShut
  {NoStop}%
\bibitem [{202(2021)}]{2021TEQUILA}%
  \BibitemOpen
  \href@noop {} {\bibfield  {journal} {\bibinfo  {journal} {Quantum Science and
  Technology}\ }\textbf {\bibinfo {volume} {6}},\ \bibinfo {pages} {024009
  (22pp)} (\bibinfo {year} {2021})}\BibitemShut {NoStop}%
\bibitem [{\citenamefont {Arute}\ \emph {et~al.}(2019)\citenamefont {Arute},
  \citenamefont {Arya}, \citenamefont {Babbush}, \citenamefont {Bacon},
  \citenamefont {Bardin}, \citenamefont {Barends}, \citenamefont {Biswas},
  \citenamefont {Boixo}, \citenamefont {Brandao}, \citenamefont {Buell},\ and\
  \citenamefont {et~al.}}]{Arute_2019}%
  \BibitemOpen
  \bibfield  {author} {\bibinfo {author} {\bibfnamefont {F.}~\bibnamefont
  {Arute}}, \bibinfo {author} {\bibfnamefont {K.}~\bibnamefont {Arya}},
  \bibinfo {author} {\bibfnamefont {R.}~\bibnamefont {Babbush}}, \bibinfo
  {author} {\bibfnamefont {D.}~\bibnamefont {Bacon}}, \bibinfo {author}
  {\bibfnamefont {J.~C.}\ \bibnamefont {Bardin}}, \bibinfo {author}
  {\bibfnamefont {R.}~\bibnamefont {Barends}}, \bibinfo {author} {\bibfnamefont
  {R.}~\bibnamefont {Biswas}}, \bibinfo {author} {\bibfnamefont
  {S.}~\bibnamefont {Boixo}}, \bibinfo {author} {\bibfnamefont {F.~G. S.~L.}\
  \bibnamefont {Brandao}}, \bibinfo {author} {\bibfnamefont {D.~A.}\
  \bibnamefont {Buell}}, \ and\ \bibinfo {author} {\bibnamefont {et~al.}},\
  }\href {\doibase 10.1038/s41586-019-1666-5} {\bibfield  {journal} {\bibinfo
  {journal} {Nature}\ }\textbf {\bibinfo {volume} {574}},\ \bibinfo {pages}
  {505–510} (\bibinfo {year} {2019})}\BibitemShut {NoStop}%
\bibitem [{\citenamefont {Vasileska}\ \emph {et~al.}(2010)\citenamefont
  {Vasileska}, \citenamefont {Goodnick},\ and\ \citenamefont
  {Klimeck}}]{2010Computational}%
  \BibitemOpen
  \bibfield  {author} {\bibinfo {author} {\bibfnamefont {D.}~\bibnamefont
  {Vasileska}}, \bibinfo {author} {\bibfnamefont {S.~M.}\ \bibnamefont
  {Goodnick}}, \ and\ \bibinfo {author} {\bibfnamefont {G.}~\bibnamefont
  {Klimeck}},\ }\href@noop {} {\emph {\bibinfo {title} {Computational
  Electronics: Semiclassical and Quantum Device Modeling and Simulation}}}\
  (\bibinfo  {publisher} {Computational electronics : semiclassical and quantum
  device modeling and simulation},\ \bibinfo {year} {2010})\BibitemShut
  {NoStop}%
\bibitem [{\citenamefont {Ajagekar}\ \emph {et~al.}(2019)\citenamefont
  {Ajagekar}, \citenamefont {Humble},\ and\ \citenamefont {You}}]{2019Quantum}%
  \BibitemOpen
  \bibfield  {author} {\bibinfo {author} {\bibfnamefont {A.}~\bibnamefont
  {Ajagekar}}, \bibinfo {author} {\bibfnamefont {T.}~\bibnamefont {Humble}}, \
  and\ \bibinfo {author} {\bibfnamefont {F.}~\bibnamefont {You}},\ }\href@noop
  {} {\bibfield  {journal} {\bibinfo  {journal} {Computers and Chemical
  Engineering}\ } (\bibinfo {year} {2019})}\BibitemShut {NoStop}%
\bibitem [{\citenamefont {Corrigan-Gibbs}\ \emph {et~al.}(2017)\citenamefont
  {Corrigan-Gibbs}, \citenamefont {Wu},\ and\ \citenamefont
  {Boneh}}]{10.1145/3102980.3102993}%
  \BibitemOpen
  \bibfield  {author} {\bibinfo {author} {\bibfnamefont {H.}~\bibnamefont
  {Corrigan-Gibbs}}, \bibinfo {author} {\bibfnamefont {D.~J.}\ \bibnamefont
  {Wu}}, \ and\ \bibinfo {author} {\bibfnamefont {D.}~\bibnamefont {Boneh}},\
  }in\ \href {\doibase 10.1145/3102980.3102993} {\emph {\bibinfo {booktitle}
  {Proceedings of the 16th Workshop on Hot Topics in Operating Systems}}},\
  \bibinfo {series and number} {HotOS '17}\ (\bibinfo  {publisher} {Association
  for Computing Machinery},\ \bibinfo {address} {New York, NY, USA},\ \bibinfo
  {year} {2017})\ p.\ \bibinfo {pages} {76–81}\BibitemShut {NoStop}%
\bibitem [{\citenamefont {Honan}\ \emph {et~al.}(2020)\citenamefont {Honan},
  \citenamefont {Lewis}, \citenamefont {Anderson},\ and\ \citenamefont
  {Cooke}}]{2020A}%
  \BibitemOpen
  \bibfield  {author} {\bibinfo {author} {\bibfnamefont {R.}~\bibnamefont
  {Honan}}, \bibinfo {author} {\bibfnamefont {T.~W.}\ \bibnamefont {Lewis}},
  \bibinfo {author} {\bibfnamefont {S.}~\bibnamefont {Anderson}}, \ and\
  \bibinfo {author} {\bibfnamefont {J.}~\bibnamefont {Cooke}},\ }\href@noop {}
  {\emph {\bibinfo {title} {A Quantum Computer Operating System}}}\ (\bibinfo
  {publisher} {Algorithms and Architectures for Parallel Processing},\ \bibinfo
  {year} {2020})\BibitemShut {NoStop}%
\bibitem [{\citenamefont {Chen}\ and\ \citenamefont {Guo}(2019)}]{2019QRunes}%
  \BibitemOpen
  \bibfield  {author} {\bibinfo {author} {\bibfnamefont {Z.~Y.}\ \bibnamefont
  {Chen}}\ and\ \bibinfo {author} {\bibfnamefont {G.~P.}\ \bibnamefont {Guo}},\
  }\href@noop {} {\  (\bibinfo {year} {2019})}\BibitemShut {NoStop}%
\bibitem [{\citenamefont {Li}\ \emph {et~al.}(2019)\citenamefont {Li},
  \citenamefont {Ding},\ and\ \citenamefont {Xie}}]{2019Tackling}%
  \BibitemOpen
  \bibfield  {author} {\bibinfo {author} {\bibfnamefont {G.}~\bibnamefont
  {Li}}, \bibinfo {author} {\bibfnamefont {Y.}~\bibnamefont {Ding}}, \ and\
  \bibinfo {author} {\bibfnamefont {Y.}~\bibnamefont {Xie}},\ }in\ \href@noop
  {} {\emph {\bibinfo {booktitle} {the Twenty-Fourth International
  Conference}}}\ (\bibinfo {year} {2019})\BibitemShut {NoStop}%
\bibitem [{\citenamefont {Murali}\ \emph {et~al.}(2019)\citenamefont {Murali},
  \citenamefont {Baker}, \citenamefont {Abhari}, \citenamefont {Chong},\ and\
  \citenamefont {Martonosi}}]{2019Noise}%
  \BibitemOpen
  \bibfield  {author} {\bibinfo {author} {\bibfnamefont {P.}~\bibnamefont
  {Murali}}, \bibinfo {author} {\bibfnamefont {J.~M.}\ \bibnamefont {Baker}},
  \bibinfo {author} {\bibfnamefont {A.~J.}\ \bibnamefont {Abhari}}, \bibinfo
  {author} {\bibfnamefont {F.~T.}\ \bibnamefont {Chong}}, \ and\ \bibinfo
  {author} {\bibfnamefont {M.}~\bibnamefont {Martonosi}}\ }(\bibinfo {year}
  {2019})\BibitemShut {NoStop}%
\bibitem [{\citenamefont {Wille}\ \emph {et~al.}(2019)\citenamefont {Wille},
  \citenamefont {Burgholzer},\ and\ \citenamefont {Zulehner}}]{2019Mapping}%
  \BibitemOpen
  \bibfield  {author} {\bibinfo {author} {\bibfnamefont {R.}~\bibnamefont
  {Wille}}, \bibinfo {author} {\bibfnamefont {L.}~\bibnamefont {Burgholzer}}, \
  and\ \bibinfo {author} {\bibfnamefont {A.}~\bibnamefont {Zulehner}},\
  }\href@noop {} {\  (\bibinfo {year} {2019})}\BibitemShut {NoStop}%
\bibitem [{\citenamefont {Bonnet}\ \emph {et~al.}(2016)\citenamefont {Bonnet},
  \citenamefont {Miltzow},\ and\ \citenamefont {Rzewski}}]{E2016Complexity}%
  \BibitemOpen
  \bibfield  {author} {\bibinfo {author} {\bibfnamefont {E.}~\bibnamefont
  {Bonnet}}, \bibinfo {author} {\bibfnamefont {T.}~\bibnamefont {Miltzow}}, \
  and\ \bibinfo {author} {\bibfnamefont {P.}~\bibnamefont {Rzewski}},\
  }\href@noop {} {\bibfield  {journal} {\bibinfo  {journal} {Algorithmica}\
  }\textbf {\bibinfo {volume} {80}},\ \bibinfo {pages} {2656} (\bibinfo {year}
  {2016})}\BibitemShut {NoStop}%
\bibitem [{\citenamefont {Kelly}\ \emph {et~al.}(2015)\citenamefont {Kelly},
  \citenamefont {Barends}, \citenamefont {Fowler}, \citenamefont {Megrant},
  \citenamefont {Jeffrey}, \citenamefont {White}, \citenamefont {Sank},
  \citenamefont {Mutus}, \citenamefont {Campbell},\ and\ \citenamefont
  {Chen}}]{2015State}%
  \BibitemOpen
  \bibfield  {author} {\bibinfo {author} {\bibfnamefont {J.}~\bibnamefont
  {Kelly}}, \bibinfo {author} {\bibfnamefont {R.}~\bibnamefont {Barends}},
  \bibinfo {author} {\bibfnamefont {A.~G.}\ \bibnamefont {Fowler}}, \bibinfo
  {author} {\bibfnamefont {A.}~\bibnamefont {Megrant}}, \bibinfo {author}
  {\bibfnamefont {E.}~\bibnamefont {Jeffrey}}, \bibinfo {author} {\bibfnamefont
  {T.~C.}\ \bibnamefont {White}}, \bibinfo {author} {\bibfnamefont
  {D.}~\bibnamefont {Sank}}, \bibinfo {author} {\bibfnamefont {J.~Y.}\
  \bibnamefont {Mutus}}, \bibinfo {author} {\bibfnamefont {B.}~\bibnamefont
  {Campbell}}, \ and\ \bibinfo {author} {\bibfnamefont {Y.}~\bibnamefont
  {Chen}},\ }\href@noop {} {\bibfield  {journal} {\bibinfo  {journal} {NATURE
  -LONDON-}\ } (\bibinfo {year} {2015})}\BibitemShut {NoStop}%
\end{thebibliography}%


%

\end{document}